\documentclass[onecolumn,preprintnumbers, sort&compress, floatfix]{revtex4}
\usepackage{amsmath}
\usepackage{amsfonts}
\usepackage{amssymb}
\usepackage{graphicx}
\usepackage{bm}
\usepackage{wrapfig}
\usepackage{caption}
\usepackage{subfigure}
\usepackage{pst-tree}

\def\be{\begin{equation}}
\def\ee{\end{equation}}
\def\bi{\bibitem}
\begin{document}

\title{Resolving the issue of branched Hamiltonian in modified Lanczos-Lovelock gravity}
\author{Soumendranath Ruz}
 \email{ruzfromju@gmail.com}
 \affiliation {Dept.of Physics, University of Kalyani, Nadia, India - 741235}

\author{Ranajit Mandal}
 \email{ranajitmandalphys@gmail.com}
 \affiliation {Dept.of Physics, University of Kalyani, Nadia, India - 741235}

\author{Subhra Debnath}
 \email{subhra_dbnth@yahoo.com}
 \affiliation {Dept.of Physics, Jangipur College, Murshidabad, India - 742213}

\author{Abhik Kumar Sanyal}
 \email{sanyal_ak@yahoo.com}
 \affiliation {Dept.of Physics, Jangipur College, Murshidabad, India - 742213}

\begin{abstract}
The Hamiltonian constraint $H_c = N{\mathcal H} = 0$, defines a diffeomorphic structure on spatial manifolds by the lapse function $N$ in general theory of relativity. However, it is not manifest in Lanczos-Lovelock gravity, since the expression for velocity in terms of the momentum is multivalued. Thus the Hamiltonian is a branch function of momentum. Here we propose an extended theory of Lanczos-Lovelock gravity to construct a unique Hamiltonian in its minisuperspace version, which results in manifest diffeomorphic invariance and canonical quantization.
\end{abstract}

\maketitle

\section{Introduction}

It is well known that gauge invariant divergences make General Theory of Relativity non-renormalizable. However, a modified fourth order gravitational action, $A = \int d^4x \sqrt{-g}[\alpha R + \beta R^2 + \gamma R_{\mu\nu}R^{\mu\nu}]$ although is renormalizable in $4$ dimension, analysis of linearized radiation reveals eight dynamical degrees of freedom, out of which $5$ massive spin-$2$ excitations are ghosts \cite{stelle}. These ghosts destroy the unitarity of the theory. Nevertheless, a particular combination of higher order curvature invariant terms produces second order field equations instead, and in the process ghosts disappear. This is known as the Gauss-Bonnet combination, ${\mathcal{G_D}} (= R^2 - 4R_{\mu\nu}R^{\mu\nu} + R_{\mu\nu\delta\sigma}R^{\mu\nu\delta\sigma})$. Unfortunately, such a combination is topologically invariant in 4 dimension. However, in $D \ge 5$, inclusion of such term leads to non-trivial dynamics and the resulting theory is known as Lanczos-Lovelock gravity \cite{love}. Lovelock invariants are second-rank symmetric tensors with vanishing covariant derivatives and depend only on the metric and its first and second derivatives. Thus, the general properties of the Einstein tensor are retained, but the theory is higher order in curvature. The Lovelock invariants are constructed in such a way that under variation, no higher than second derivatives appear in the field equations, supposedly avoiding the problem of unitarity. Apart from normalization factor, one can express lovelock invariants $\mathcal L_m$ as,
\be \mathcal L_m = \delta^{a_1b_1...a_mb_m}_{c_1d_1...c_md_m}R_{a_1b_1}^{c_1d_1}.....R_{a_mb_m}^{c_md_m} \ee
where $\delta^{a_1b_1...a_mb_m}_{c_1d_1...c_md_m}$ is the totally antisymmetric product of Kronecker deltas and $R_{a_mb_m}^{c_md_m}$
is the $D$-dimensional curvature tensor. For a space with an even number of dimensions, $D = 2m$, the Lovelock invariant is a topological one and leads to a total derivative term. All the higher order invariants, $D = 2m-1$, can then be shown to vanish identically using symmetry properties of curvature tensor  \cite{Rizzo}. On the other hand, for $D = 2m + 1$, $\mathcal L_m$ are true dynamical objects normally associated with the Einstein-Hilbert term. Therefore a large number of Lovelock invariants exists for arbitrary dimension. Interestingly, the Lovelock invariants are found to be just the forms admissible by the higher order curvature terms generated in perturbative critical string theory \cite{string}. Therefore, from physical consideration the number of Lovelock invariants may be restricted. For example, in string-motivated higher dimensional theory, viz. in $10$ dimensional superstring gravity, only three additional terms, other than the Einstein-Hilbert term appear \cite{ADD}. For the present purpose, we consider only the Gauss-Bonnet term, for which the Lanczos-Lovelock action in $D$ dimension reads,
\be\label{A3} A_1=\int\sqrt{-g}\;d^Dx\left[\frac{R - 2\Lambda_0}{16\pi G_{\mathrm N}}+\gamma\mathcal{G_D} \right]+ \Sigma_{R_D} + \Sigma_{\mathcal{G_D}},\ee
where, $G_{\mathrm N}$ is the Newton's gravitational constant, $\gamma$ being the coupling constant and
\begin{subequations}\label{2.5}
\begin{align}
&\Sigma_{R_D} = {1\over 8\pi G_{\mathrm N}}\oint_{\partial\mathcal{V}} d^{D-1}x \sqrt{h} K, \label{2.5a}\\
&\Sigma_{\mathcal{G_D}} = 4\gamma\oint_{\partial\mathcal{V}}\sqrt{h}d^{D-1} x\left( 2G_{ij}K^{ij} + \frac{\mathcal{K}}{3}\right) \label{2.5b}
\end{align}
\end{subequations}
are the supplementary boundary terms corresponding to Ricci scalar $R$ and Gauss-Bonnet term $\mathcal{G_D}$ respectively in $D$ dimension \cite{others}. In the expression (\ref{2.5b}) the symbol $\mathcal{K}$ stands for $\mathcal{K}=\left( K^3 - 3K K^{ij}K_{ij} + 2K^{ij}K_{ik}K^k_j \right)$  where, $K$ is the trace of the extrinsic curvature tensor $K_{ij}$, and $G_{ij}$ is the Einstein tensor built out of the induced metric $h_{ij}$ on the boundary. Very importance of such an action in the context of astrophysics and cosmology has already been established. For example, asymptotically dS/AdS together with flat solutions \cite{3.1} and conformal anomaly from higher derivative gravity in AdS/CFT correspondence \cite{3.2} have been realized. For the zero-temperature background, solutions both in pure Gauss-Bonnet gravity and that with non-trivial matter have been found in asymptotically Lifshitz spacetimes in five dimensions \cite{3.3}. Static solution corresponding to such an action in vacuum has been presented \cite{3.4} and Birkhoffs theorem has been established \cite{deser}. Exact topological Black-hole solutions and thermodynamic properties of Black-hole horizon have been studied extensively \cite{cai}. It has also been studied in the context of steep inflationary scenario \cite{3.5} and has been found to admit Friedmannn-like classical solutions \cite{3.6}. Bouncing cosmological models represent the most promising alternative theory to the inflationary paradigm, since it provides theoretical values of the cosmological parameters, that fit well \cite{Eliz} with recent Planck data \cite{Planck} Recently, it has been shown that some of the Loop Quantum Cosmology corrected Gauss-Bonnet-modified gravity theories can successfully realize exponential or power law bouncing cosmological solutions \cite{Haro}. Successful dark energy models with Gauss-Bonnet-Dilatonic coupled action, required to explain late-time cosmological acceleration also exist in the literature \cite{Noj}\\

However, while performing canonical analysis of Lovelock action under $4+1$ decomposition, Deser and Franklin noticed that the presence of cubic kinetic terms and quadratic constraints make the theory intrinsically nonlinear \cite{deser}. Even its linearized version is cubic rather than quadratic. Such a pronounced exotic behaviour of the action does not allow Hamiltonian formulation of Lanczos-Lovelock gravity following conventional Legendre transformation. As a result, diffeomorphic invariance is not manifest and standard canonical formulation of the theory is not possible. Such a situation arises because the Lagrangian is quartic in velocities and as a result, the expression for velocities are multivalued functions of momentum, resulting in the so called multiply branched Hamiltonian with cusps. This makes classical solution unpredictable as at any time one can jump from one branch of the Hamiltonian to the other. Further, the momentum does not provide a complete set of commuting observable resulting in non-unitary time evolution of quantum states. Although, Lanczos-Lovelock shows unitary time evolution of quantum  states, when expanded perturbatively about the flat Minkowski background; non-perturbatively, the situation is miserable. Thus the main aim of constructing Lanczos-Lovelock gravity falls short non-perturbatively. Further, it shows yet another pathology by spontaneous breaking the time translational symmetry of the theory. Note that, so far there exists only a handful of techniques to resolve the issue of branched Hamiltonian, and that too have been made in some toy models; and for gravity, in minisuperspace models only. There is no standard and unique theory to resolve the issue, and therefore handling the problem for the general Lanczos-Lovelock action is awesome. The present aim here is to explore the associated problem in the Robertson-Walker(R-W) minisuperspace and to show that an additional scalar curvature squared term ($R^2$) term alleviates the problem of branching, restoring time translational symmetry. Note that, the removal of all the pathological behaviour of Lanczos-Lovelock gravity although is generic for any higher order curvature invariant term, the reason for adding scalar curvature squared term in particular is straightforward. When expanded perturbatively about Minkowski background, $5$ massive spin-$2$ particles appear due to the presence of $R_{\mu\nu}R^{\mu\nu}$ term in the action, which are ghosts. However neither fourth order gravity in general, nor the presence of $R^2$ term in the action is responsible for the appearance of ghosts. It is therefore safe to add $R^2$ term to establish unitarity, both perturbatively and non-perturbatively. Recently, canonical formulation of Weyl tensor in arbitrary dimension has been performed in the whole superspace \cite{Deruelle}. Weyl tensor squared term may be expressed in terms of Gauss-Bonnet ($\mathcal{G_D}$) term as

\be C_{\alpha\beta\mu\nu}C^{\alpha\beta\mu\nu} = \mathcal{G_D} + 4\left({D-3\over D-2}\right)R_{\mu\nu}R^{\mu\nu} - {D(D-3)\over (D-1)(D-2)}R^2\ee
However, such canonical formulation establishes diffeomorphic invariance and does not encounter the pathology of branching due to the appearance of Gauss-Bonnet term. So, the removal of the pathologies of Lanczos-Lovelock gravity is not due to suppression of large number of degrees of freedom in the minisuperspace model under consideration, rather it is also generic. The reason for considering minisuperspace is the following. In the context of quantum cosmology there is no time evolution, since $H|\psi> = 0$. Therefore, there is no scope to test the issue of unitarity. However, in the minisuperspace model, due to the presence of higher order term in the action, the quantum version takes Schr\"odinger-like form, where an internal parameter acts as the time parameter. Therefore, in view of the hermitian nature of the effective Hamiltonian, unitarity of the modified Lanczos-Lovelock gravity may be established.\\

In the following section, we expatiate the problem associated with canonical formulation of the action (\ref{A3}). In section III, we briefly enunciate recent couple of attempts in this regard, following Legendre-Fenchel transformation \cite{Chi} and generalized Legendre transformation \cite{avra}, to show that the resulting Hamiltonians are not related through canonical transformation. Further, both the Hamiltonians show spontaneous breaking of time translational symmetry, as well. Since there is no scope to study the behaviour of the corresponding quantum theories under some appropriate semi-classical approximation, therefore there is no way in principle, to choose the correct formalism in this regard. In section IV, we take up extended Lanczos-Lovelock gravity, cast the action in canonical form, establish diffeomorphic invariance and follow standard canonical quantization scheme. We then make semiclassical approximation to establish oscillatory behaviour of the wave-function around classical trajectory. In section V, we discuss the issue of spontaneous breaking and making of time translational symmetry. In the process, all the problems associated with Lanczos-Lovelock gravity have been alleviated. Some concluding remarks are made in section VI.

\section{Problem associated with Canonical formulation}
In the $D$ dimensional Robertson-Walker minisuperspace
\be\label{m5}\begin{split}
ds^2 &= -N^2 dt^2 + a(t)^2\Bigg[{dr^2\over 1-kr^2} + r^2d\Omega^2 + d X_{\delta}^2\Bigg],
\end{split}\ee
\noindent
where $\delta = (D-4)$ stands for extra dimension and $d\Omega^2 = (d\theta^2 + sin^2\theta d\phi^2) $, the expressions for the Ricci scalar ($R$) and the Gauss-Bonnet term ($\mathcal{G_D}$) are
\begin{subequations} \begin{align}
R &= \frac{D-1}{N^2}\left[ \frac{2\ddot a}{a} + \frac{(D-2)\dot a^2}{a^2} - \frac{2\dot a\dot N}{aN} \right] + \frac{6k}{a^2} \label{r}\\
\mathcal{G}_D &= \frac{(D-3)}{a^3N^2}\Bigg[(D-4)\frac{\dot a^2}{a}\left(\Delta_1\frac{\dot a^2}{N^2} + 12k\right) + 4\Delta_1\frac{\dot a^2}{N^2}\Bigg(\ddot a - \frac{\dot a\dot N}{N}\Bigg)+ 24k\left(\ddot a - \frac{\dot a\dot N}{N}\right)\Bigg] \label{g}
\end{align}\end{subequations}
\noindent
with $\Delta_1 = (D-1)(D-2)$. Plugging in the above expressions in action (\ref{A3}) and after cancelling the total derivative terms appearing under integration by parts with the supplementary boundary terms as usual, it reads
\be\label{1}\begin{split}
A_1 &= \int\Bigg[\frac {a^{(D-3)}}{\kappa}\left(3kN - \Delta_1\frac{\dot a^2}{2N} - Na^2\Lambda_0 \right) - \gamma\Delta_2\frac{\dot a^2}{N}a^{(D-5)}\left( \Delta_1\frac{\dot a^2}{3N^2} + 12k \right)\Bigg]dt
\end{split}\ee
where, $\Delta_2 = (D-3)(D-4)$ and $\kappa = 8\pi G_\mathrm{N}$. Therefore the canonical momenta are
\begin{subequations} \begin{align}
& p_N = 0 \\
& p_a = - \frac{\Delta_1\dot a a^{(D-3)}}{\kappa N} - \frac{4\gamma\Delta_2\dot a}{Na^{(5-D)}} \left( \frac{\Delta_1\dot a^2}{3N^2} + 6k \right) \label{pa1}
\end{align}\end{subequations}
Corresponding Hamiltonian, obtained from $N$ variation equation is
\be\begin{split}
H_c &= N \Bigg[-\frac {a^{(D-3)}}{\kappa}\left( \frac{\Delta_1}{2}\frac{\dot a^2}{N^2} + 3k - a^2\Lambda_0 \right) - \gamma \Delta_2\frac{\dot a^2 a^{(D-5)}}{N^2}\left(\Delta_1\frac{\dot a^2}{N^2} + 12k \right)\Bigg]
\end{split}\ee
Although it is unique in terms of the velocities, there indeed exists three different Hamiltonians in terms of the phase space variables, since the momentum (\ref{pa1}) appears in third degree algebraic equation in $\dot a$, and one can have either one or three real values of $\dot a$ corresponding
to a given value of momentum $p_a$. This problem was noticed by Deser and Franklin, who stated it as ``a most un-Hamiltonian system" \cite{deser}. Such multi-valued Hamiltonian with cusps, usually called the branched Hamiltonian, makes the classical theory unpredictable and does not allow standard canonical
formulation of the theory. Further, since energy has to be an observable, the momentum does not ensure a complete set of commuting observable. Also, the lowest energy solution of the system "spontaneously breaks time translation", because at the cusps the velocity is non-vanishing. Finally, the theory also suffers from the disease of having non-unitary time evolution of the quantum state. Last point is most important in the context of Lanczos-Lovelock gravity. Expanding Lanczos-Lovelock action in the perturbative series about the linearized theory reveals that, it is free from ghosts and thus is unitary. However, non-perturbatively the theory lacks unitary time evolution, as mentioned. This demonstrates that perturbative analysis is misleading. Such unpleasant issue arising out of branched Hamiltonian was addressed long ago \cite{3.7}. Starting from a toy model, Henneaux, Teitelboim and Zanelli \cite{3.7} had shown that, in the path integral formalism one can associate a perfectly smooth quantum theory which possesses a clear operator interpretation and a smooth, deterministic, classical limit. Nevertheless, it puts up question on the standard classical variational principle and of course on the canonical quantization scheme. The same issue has also been addressed by several authors in the recent years \cite{3.8, 3.9, 3.10}. However, in order to solve the problem they also had to tinker with some fundamental aspects, e.g., loosing Heaviside function to obtain manifestly hermitian convolution \cite{3.8}, sacrificing the Darboux coordinate to parametrize the phase space \cite{3.9} and the usual Heisenberg commutation relations \cite{3.10}. Therefore none of these techniques is fully developed or rigorous. The latest development in the path integral formalism has been presented by Chi and He \cite{Chi}. They \cite{Chi} proposed the Legendre-Fenchel Transformation (LFT) method to obtain single-valued Hamiltonian in a toy model, considered earlier in \cite{3.7, 3.8}, which is suitable for studying the ground state wave-function. This may be applied to path integral approach for quantization. Legendre-Fenchel transformation (LFT) is applicable to non-convex Lagrangians and it reduces to the conventional Legendre transformation for convex Lagrangians. More recently, Avraham and Brustein \cite{avra} have developed a modified version of Dirac's constrained analysis \cite{dirac, Dirac} following Generalized Legendre Transformation (GLT), which is also of interest. However different techniques lead to different Hamiltonian corresponding to the same action, and these Hamiltonians are not related through canonical transformation, as we demonstrate in section III. Further, quantization of these Hamiltonians is also an awesome task, and therefore there is no way to find a classical limit in order to pick up the correct technique. In this connection, we adopt a completely different scheme to alleviate the problem, by associating an additional scalar curvature invariant term ($R^2$) in the Lanczos-Lovelock action and in the process, we propose an extended Lanczos-Lovelock theory of gravity. Note that ($R^2$) term plays a crucial role in the early universe. For example, the dominance of such term leads to inflation without invoking phase transition \cite{staro} and canonical quantization yields a Schr\"odinger like equation, leading to quantum mechanical probability interpretation in a straight forward manner \cite{aks1, aks3}. Additionally, semiclassical wave-function obtained under WKB approximation has been found to be oscillatory, indicating that the region is classically allowed and it is strongly peaked about a set of solutions to the classical field equations \cite{aks4, aks5}. In the following section, we briefly demonstrate the LFT model followed by Chi and He \cite{Chi} in the context of latest development along path-integral quantization scheme. However, since our interest is in canonical quantization, we therefore also demonstrate the GLT scheme \cite{avra} following constraint analysis developed by Dirac \cite{dirac, Dirac}. We then compare the Hamiltonians produced by LFT and GLT for the same toy model to show that they are not related through canonical transformation.

\section{Attempts to handle the issue of branched Hamiltonian}

\subsection{Legendre-Fenchel transformation}
In a recent work, Chi and He \cite{Chi} addresses the issue of  branched Hamiltonian and proposed construction of a single-valued Hamiltonian out of a non-convex Lagrangian, by applying Legendre-Fenchel transformation. For the purpose of demonstration, they considered a toy model in the form,
\be\label{chL} L = \frac{1}{4}{\dot\phi}^4 - \frac{1}{2}{\dot\phi}^2.\ee
The conjugate momentum corresponding to the above non-convex Lagrangian is
\be\label{chp} p = \frac{\partial L}{\partial \dot\phi} = \dot\phi^3 - \dot \phi,\ee
and the conventional Legendre transformation yields,
\be H = \frac{3}{4}{\dot\phi}^4 - \frac{1}{2}{\dot\phi}^2,\ee
which is clearly multi-valued function in canonically conjugate momentum $p$, since each given $p$ corresponds to one or three values of $\dot\phi$. However, they handled this situation via Legendre-Fenchel transformation $H(p) = {\mathrm{Sup}}_{\dot\phi\in N(p)}[p\dot\phi - L(\dot\phi)]$, where $N(p) = {\dot\phi}|{\mathrm{Sup}~{p(\dot\phi)-L(\dot\phi)<\infty}}$ ensures finiteness of the Hamiltonian (Sup stands for Supremum). The Hamiltonian obtained from Legendre-Fenchel transformation equals the minimal intercept (with an overall sign flip) of a line with slope $p$. Since momentum in the case under consideration is conserved, one can solve $\dot\phi$ equation algebraically to obtain three values of $\dot \phi$ as functions of $p$. Further, because the mimima and the maxima of momentum are fixed at $p_1=-{2\over 3\sqrt 3}$ and $p_2={2\over 3\sqrt 3}$ respectively, the unique right and the left tangent points of $p~ \mathrm{versus}~\dot\phi$ curve give the minima for $p\in\left(\frac{2}{3\sqrt 3}, +\infty\right)$ and $p\in\left(-\infty,-\frac{2}{3\sqrt 3}\right)$ respectively. The expression for the tangent points ${\dot\phi_1}(p)$ and $\dot\phi_3 (p)$ are
\be\label{chphi}\begin{split}
& {\dot\phi_1}(p) = \frac{\left(\frac{2}{3}\right)^{\frac{1}{3}}}{f(p)} + \frac{f(p)}{2^\frac{1}{3} 3^\frac{2}{3}}\ \ \ \text{and} \ \ \ {\dot\phi_3}(p) = -\frac{1 + i \sqrt{3}}{2^{2\over 3} 3^{1\over 3} f(p)}- \frac{(1 - i\sqrt{3})f(p)}{2^{4\over 3} 3^{2\over 3}}.
\end{split}\ee
where $f(p) = \left(9 p +  \sqrt{3( 27 p^2-4)}\right)^{1\over 3}$. Now, in view of the expression for momentum, the Lagrangian may finally be expressed as
\be\begin{split}
&L(\dot\phi_1) = \frac{1}{4}{\dot\phi_1}^4 - \frac{1}{2}{\dot\phi_1}^2 = \frac{1}{4}p{\dot\phi_1} - \frac{1}{4}{\dot\phi_1}^2 \ \ \text{and} \ \ L(\dot\phi_3) = \frac{1}{4}{\dot\phi_3}^4 - \frac{1}{2}{\dot\phi_3}^2
=\frac{1}{4}p{\dot\phi_3} - \frac{1}{4}{\dot\phi_3}^2\end{split}\ee
Since, for $p=0$, the right and left tangent points give the same minimal intercept, so the single-valued Hamiltonian is found to take the form
\be\begin{split}& H_1(p) = p \dot \phi_1(p) - L(\dot \phi_1(p)),~~\text{for}~~p\in(0,+\infty)\\& H_2(p) = p \dot \phi_3(p) - L(\dot \phi_3(p)),~~\text{for}~~p\in(-\infty,0)\end{split}\ee
Explicit form of the Hamiltonian for the two regions therefore are
\be\begin{split}\label{chh} &H_1 = \frac{3}{4}p\dot\phi_1 + \frac{1}{4}{\dot\phi_1}^2= \frac{1}{4}\left[\frac{\left(\frac{2}{3}\right)^{\frac{1}{3}}}{f(p)} + \frac{f(p)}{2^\frac{1}{3} 3^\frac{2}{3}}\right]\left[\frac{\left(\frac{2}{3}\right)^{\frac{1}{3}}}{f(p)} + \frac{f(p)}{2^\frac{1}{3} 3^\frac{2}{3}}+3p\right]\\
&H_2 = \frac{3}{4}p\dot\phi_3 + \frac{1}{4}{\dot\phi_3}^2={1\over 4}\Bigg[\frac{1 + i \sqrt{3}}{2^{2\over 3} 3^{1\over 3} f(p)}
+\frac{(1 - i\sqrt{3})f(p)}{2^{4\over 3} 3^{2\over 3}}\Bigg]\Bigg[\frac{1 + i \sqrt{3}}{2^{2\over 3} 3^{1\over 3} f(p)}
+\frac{(1 - i\sqrt{3})f(p)}{2^{4\over 3} 3^{2\over 3}}-3p\Bigg]
\end{split}\ee
The $H ~\mathrm{versus}~p$ plots of the two regions are glued to obtain a unique and smooth Hamiltonian as shown in figure 1. Both the Hamiltonians produce correct Euler-Lagrange equation of the system under consideration. Further, using the above single-valued Hamiltonian Chi and He \cite{Chi} found the vacuum state which is useful for path integral quantization. However, the single valued Hamiltonian thus obtained, is not suitable for canonical quantization, and therefore there is no way to check if appropriate classical behaviour is retrievable under some semiclassical approximation. Although the method is mathematically rigorous, its applicability in a more realistic cosmological model, eg. Lanczos-Lovelock gravity, is mathematically very complicated, if not impossible. Next, we therefore briefly review the attempt to construct a single valued Hamiltonian out of Lanczos-Lovelock gravity (\ref{A3}) following GLT \cite{avra}.

\begin{figure}
\begin{center}
\includegraphics[height = 2.0 in, width = 3.0 in]{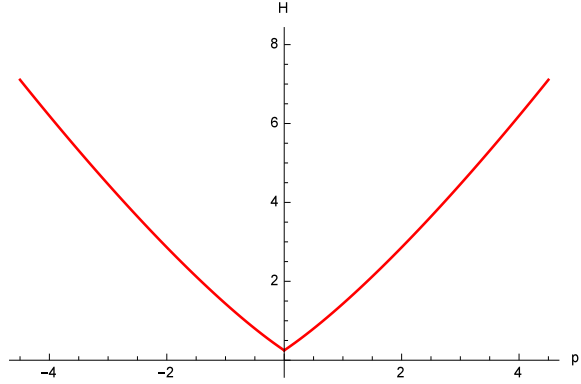}
\end{center}
\caption{The figure reproduces the Hamiltonian versus momentum plot of Chi and He (\cite{Chi}), which was obtained following Legendre-Fenchel transformation corresponding to the toy model (\ref{chL}). The Hamiltonian is smooth and has a non-zero minima.}
\end{figure}

\subsection{Generalized Legendre Transformation}

In a recent article, Avraham and Brustein \cite{avra} handled the issue of branching for Lanczos-Lovelock gravity with zero spatial curvature ($k = 0$), adopting generalized Legendre transformation. In this technique, first the phase space is extended by adding new variable and then constraints are imposed to remove the additional variable. The phase space coordinates should be so defined that the constraints are solved and the number of variables are reduced to the original number of degrees of freedom. The end result is a single valued Hamiltonian which is compatible with the symplectic structure on phase space. Briefly we attempt to expatiate the technique in connection with the action (\ref{A3}) for non-zero spatial curvature ($k\ne 0$), which reduces to action (\ref{1}) in the R-W minisuperspace model. For the sake of simplicity, let us take $D = 5$ dimensional space-time. The action (\ref{1}) therefore reduces to

\be\label{AA4} A_2 = \int \left[\frac{a^2}{\kappa}\left(-6\dot a^2 + 3k - \Lambda_0 a^2\right) - 8\gamma\dot a^2\left(\dot a^2 + 3k\right)\right]dt.\ee
Under a judicious choice of the variable viz., $a=e^x$, the corresponding point Lagrangian is expressed as,
\be \label{l1} L = 3k\left(\frac{1}{\kappa} - 8\gamma\dot x^2 \right)e^{2x}-\left(8\gamma\dot x^4 + \frac{6\dot x^2+\Lambda_0}{\kappa} \right)e^{4x}\ee
So the momentum canonical to $x$ is
\be p_x = -48k\gamma\dot xe^{2x} - 32\gamma\dot x^3e^{4x} - \frac{12}{\kappa}\dot xe^{4x} \ee
Now choosing $\dot x = Q$ as an independent variable, the Lagrangian (\ref{l1}) becomes degenerate. Therefore, one can introduce a Lagrangian multiplier $\lambda$ as
\be\begin{split}
\tilde{L} &= 3k\left(\frac{1}{\kappa} - 8\gamma Q^2 \right)e^{2x}-\left(8\gamma Q^4 + \frac{6Q^2+\Lambda_0}{\kappa} \right)e^{4x} + \lambda(\dot x-Q)
\end{split}\ee
and follow Dirac technique of constrained analysis, to find $\lambda= p_x = -48k\gamma Qe^{2x} - 32\gamma Q^3e^{4x} - \frac{12}{\kappa}Qe^{4x}$. So canonical Hamiltonian becomes,
\be H = -3k\left(8\gamma Q^2 + \frac{1}{\kappa} \right)e^{2x} + \left(\frac{\Lambda_0}{\kappa} - 24\gamma Q^4 - \frac{6}{\kappa}Q^2 \right)e^{4x} \ee
Now solving generalized Legendre equation in terms of the new functions $f(x, Q)$ and $g(x, Q)$, such that $\{f,g\}_D = 1$, where subscript $D$ stands for Dirac bracket \cite{avra} one obtains,
\be\begin{split}
&\partial_x f\partial_Q g - \partial_Q f\partial_x g = \partial^2_Q L(x,Q) = -48k\gamma e^{2x} - 96\gamma Q^2e^{4x} - \frac{12}{\kappa}e^{4x} \end{split}\ee
Finally, choosing $g=Q$, one can find $f$ in the following form,
\be f = -24k\gamma e^{2x} - 24\gamma Q^2e^{4x} - \frac{3}{\kappa}e^{4x} \ee
From the above form of $f$, it is apparent that unless $k=0$, it is not possible to express $H(x,Q)$ simply in terms of $f$ and $g$. So choosing $k=0$, the canonical Hamiltonian is found in the following form,
\be H(f, g) = \left(24\gamma g^4 + \frac{6}{\kappa}g^2 - \frac{\Lambda_0}{\kappa} \right)\frac{f}{24\gamma g^2 + \frac{3}{\kappa}}, \ee
which in the absence of cosmological constant $\Lambda_0$, reduces to
\be \label{ham}H(f, g) = fg^2 \left[1+ {1\over 8\gamma\kappa g^2 + 1}\right],\ee
and was obtained by Avraham and Brustein \cite{avra}. It is of-course true that Hamilton's equations produce correct classical field equations and the minima of the above Hamiltonian gives
\be\label{cl} a = a_0 e^{\pm \sqrt{1\over\alpha} t},\ee
where, $\alpha = - 4\kappa\gamma$, leading to expanding or contracting universe. This is nothing special, since it is just the solution to the classical field equation, which simply proves that the Hamiltonian arrived at, produces correct classical field equations. The question -``is the Hamiltonian arrived at, produces a viable quantum description?" - may only be answered if under some suitable semiclassical approximation, the wave function shows oscillatory behaviour about the above classical solution (\ref{cl}). In gravitation, reparametrization invariance, constraints the Hamiltonian $H(f,g)$ to vanish, and therefore treating $f$ as a variable and $g$ as its canonical momentum as depicted through Dirac bracket, the quantum description of such a Hamiltonian (\ref{ham}) reads,
\be \label{Q}(4\gamma\kappa \hat g^4 + \hat g^2)\psi = 0. \ee
The above quantum equation looks deceptively simple, as Cauchy data doesn't match boundary data. The reason is, the quantum version of the second order classical field equation turns out to be fourth order. To solve the Euler-Lagrange equation, one requires two initial data ($a,~\dot a$), which matched two boundary data ($\delta a =0$ at the two boundaries). However, the quantum version of the same theory requires four initial data and so equation (\ref{Q}) does not appear to produce a viable quantum description of the theory under consideration. Further, we have also observed that such Hamiltonian formulation is possible only under the choice $k = 0$, i.e., assuming universe to be flat a-priori, and of-course without matter field. The so-called $(fgh)$ model of Zhao, Yu and Xu \cite{3.9} may be translated to the Lanczos-Lovelock action (\ref{AA4}), under the condition $f > 0, g < 0$, which when translated, gives the Gauss-Bonnet coupling $\gamma < 0$ and the scale factor $a^2 > 24\alpha^2\kappa k$. The later is true in general, only for $k = 0$ again. Therefore attempt to construct a single-valued Hamiltonian out of Lanczos-Lovelock gravity for arbitrary curvature parameter is still obscure. \\

\noindent
One can also notice that the first example treated by Avraham and Brustein \cite{avra} taking the toy model
\be\label{lag} L = {1\over 4} \dot x^4 - {k\over 2} \dot x^2 - {1\over 2}w x^2\ee
finally produces a Hamiltonian in the form
\be \label{H}H = {w f^2 \over 2(3g^2-k)^2} + {3\over 4}g^4 -{k\over 2} g^2,\ee
where,
\be g = Q = \dot x,~\mathrm{and}~f = x(3Q^2 - k) =  x(3\dot x^2 - k).\ee
It is important to note that, the Hamilton's equations of motion under consideration are ($w = 0$)
\be \dot g = - {\partial H\over\partial f} = 0;~\dot f = {\partial H\over\partial g} = 3 g^3 - k g \Longrightarrow \dot x^2 = C_1.\ee
Clearly, it does not produce Euler-Lagrange equation of motion
\be \dot x^3  - k\dot x = C_2.\ee
Additionally, one can observe also that for $w = 0,\; \mathrm{and}\;k = 1$, the Lagrangian (\ref{lag}) is identical to (\ref{chL}). The Hamiltonian in the present case
\be\label{Ham} H = {3\over 4}g^4 -{1\over 2} g^2,\ee
has also been obtained by, Zhao, Yu and Xu \cite{3.9} following a slightly different route, but sacrificing Darboux co-ordinate as well. This Hamiltonian has been plotted against the momentum in figure 2, and is found to be potentially different from the one shown in figure 1. Of-course, the Hamiltonian (\ref{Ham}) does not transform to (\ref{chh}) and vice-versa, under any canonical transformation. Therefore, it is clear that the issue of branched Hamiltonian has not been resolved uniquely, as yet.
\begin{figure}
\begin{center}
\includegraphics[height = 2.0 in, width = 3.0 in]{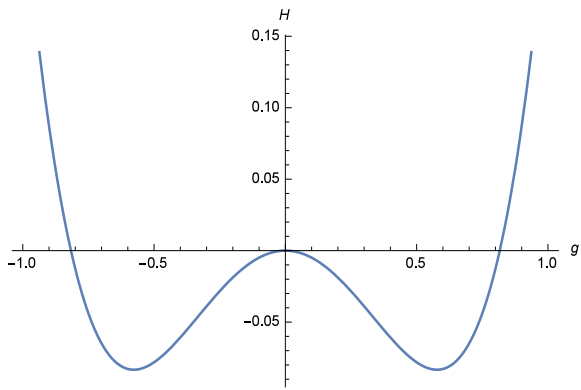}
\end{center}
\caption{The Hamiltonian versus momentum plot for the same toy model (\ref{chL}) obtained following generalized Legendre transformation shows two minima, which in potentially different from figure 1.}
\end{figure}

\section{Constructing a unique phase-space Hamiltonian}\label{abhik}
At this end, it is apparent that despite distinctive efforts in recent years, particularly by Chi and He \cite{Chi} and Avraham and Brustein \cite{avra}, the issue of branched Hamiltonian is far from being resolved. It is also important to mention that in the process of obtaining a single-valued Hamiltonian, translational symmetry is spontaneously broken (see section V) in all the three different attempts \cite{Chi, 3.10, avra}. Further, all the Hamiltonians are different and are not related via canonical transformation. Finally, attempt to construct a single valued Hamiltonian for Lanczos-Lovelock gravity with non-zero spatial curvature ($k \ne 0$) is still in vain. Under this circumstances, we observe \cite{can} that the problem is alleviated, if higher degree terms are associated with higher order ones in the action. In fact, in most of the physical situations, particularly in gravity, they appear together. As for example, all higher order curvature invariants $R^2$, $R_{\mu\nu}R^{\mu\nu}$, $R_{\mu\nu\gamma\delta}R^{\mu\nu\gamma\delta}$ etc. contain both the terms with higher order and higher degree. Canonical formulation of gravity with such terms never showed pathological behaviour such as branched Hamiltonian in minisuperspace models \cite{aks1, aks3, aks4, aks5}. Let us therefore supplement action (\ref{A3}) by $R^2$ term and express the complete action as
\be\label{A4}\begin{split}
A_3 &= \int\sqrt{-g}\;d^Dx\left(\frac{R - 2\Lambda_0}{2\kappa} + \beta R^2 + \gamma\mathcal{G}_D  \right) + \Sigma_{R_D} + \Sigma_{{R^2}_{D}} + \Sigma_{\mathcal{G}_D},
\end{split}\ee
where, $\Sigma_{{R^2}_{D}} = 4\beta\oint_{\partial\mathcal{V}} RK\sqrt{h} d^{(D-1)} x $ is the boundary term required to supplement the gravitational action carrying a scalar curvature squared term $R^2$ in $D$-dimension. Before we proceed, let us mention that due to reparametrization invariance, the gravitational Hamiltonian is constrained to vanish. As a result, the Wheeler-deWitt equation is devoid of time evolution and standard probabilistic interpretation. In fact, the quantized version of the higher order gravitational action
\be A_4 = \int\left[A C_{\mu\nu\gamma\delta} C^{\mu\nu\gamma\delta} + B(R - 4\Lambda)^2\right]\sqrt{-g}~d^4x\ee
presented by Boulware \cite{Boul} is also not free from the same disease. In this regard, Horowitz \cite{Horowitz} proposed a scheme to obtain a Schr\"odinger like equation corresponding to higher order theory of gravity, in the Robertson-Walker minisuperspace model. For this purpose, he started with an auxiliary variable by varying the action with respect to highest derivative appearing in the action and expressed the action in the canonical form. After phase-space formulation, he replaced the auxiliary variable appearing in the Hamiltonian by the basic variable, viz., the extrinsic curvature tensor $K_{ij}$. However, the whole formulation was performed in flat ($k = 0$) space. In order to take care of the boundary terms associated with the action for $k\ne 0$, the technique was further developed by Sanyal and his co-workers \cite{aks4, aks5} in the Robertson-Walker minisuperspace model, along the following prescription,
\begin{itemize}
  \item Split the boundary term for $R^2$ into two parts, viz.
  \be\label{2.6}\begin{split}
  \Sigma_{R^2_{D}} &= \Sigma_{R^2_{D_1}}+\Sigma_{R^2_{D_2}} = 4\beta\int \left[{^{(D-1)}R}+ ({^DR}-{^{(D-1)}R}) \right] K\sqrt{h} d^{(D-1)} x,
  \end{split}\ee
  where ${^{(D-1)}R}$ is the Ricci scalar built out of $h_{ij}$ and ${^DR}$ is the usual Ricci scalar in $D$ dimension.
  \item Express the action in terms of $h_{ij}$ and remove total derivative terms under integration by parts, which cancel $\Sigma_{R_D}$, $\Sigma_{R^2_{D_1}}$ and $\Sigma_{\mathcal{G}_D}$.
  \item Introduce auxiliary variable in the action following Horowitz's proposal \cite{Horowitz}. Integration by parts then takes care of the boundary term $\Sigma_{R^2_{D_2}}$. The action is then automatically expressed in canonical form.
  \item Find Hamiltonian constraint equation from $N$ variation equation which guarantees its diffeomorphic invariance. Now expressing it in terms of the basic variables ($h_{ij}, K_{ij}$), standard prescription for canonical quantization follows.
\end{itemize}
The first and second points appear from the fact that the auxiliary variable should only be introduced in those particular terms, which vanish at flat space. This was suggested by Boulware \cite{Boul}, not to treat linear gravity (Einstein-Hilbert action) as higher order theory. However, it was noticed \cite{aks1, aks3, aks4, aks5} that $R^2$ action also includes a term ($k \ddot a$ in Robertson-Walker metric) which vanishes at flat space. It should be integrated out by parts before the introduction of auxiliary variable. Auxiliary variable is introduced, as mentioned in the third point, so that it is not required to fix $K_{ij}$ at the boundary. Since, $K_{ij}$ is treated as basic variable, so if it is kept fixed at the boundary, the action must also be varied with respect to $K_{ij}$. In that case, classical solution is restricted by and large. Note that the action may be varied with respect to the auxiliary variable to obtain its definition only. Finally, the auxiliary variable should be replaced by basic variable $K_{ij}$, which is the requirement of the fourth point. Here, we follow the same route, i.e. recast the action under the choice $z = h_{ij} = a^2$, integrate it by parts to cancel the total derivative terms with the boundary terms $\Sigma_{R_D}$, $\Sigma_{R^2_{D_1}}$ and $\Sigma_{\mathcal{G}_D}$, retaining only $\Sigma_{{R^2}_{D_2}}$. In the process the action (\ref{A4}) reads
\be\label{38}\begin{split}
A_3 &= \int\bigg[\frac{1}{\kappa}\left(3kNz^{\frac{\mathcal{D}_3}{2}} - \frac{\Delta_1}{8}\frac{z^{\frac{\mathcal{D}_5}{2}}\dot z^2}{N} - \Lambda_0Nz^{\frac{\mathcal{D}_1}{2}} \right) - \gamma\bigg(\frac{\Delta_1\Delta_2}{48}\frac{z^{\frac{\mathcal{D}_9}{2}}\dot z^4}{N^3} + 3k\Delta_2\frac{z^{\frac{\mathcal{D}_7}{2}}\dot z^2}{N} \bigg) \\
&+ \beta\bigg\{ \mathcal{D}_1^2\bigg( \frac{z^{\frac{\mathcal{D}_5}{2}}\ddot z^2}{N^3} + \frac{z^{\frac{\mathcal{D}_5}{2}}\dot z^2\dot N^2}{N^5} - 2\frac{z^{\frac{\mathcal{D}_5}{2}}\dot z\ddot z\dot N}{N^4} + \frac{\mathcal{D}_4}{2}\frac{z^{\frac{\mathcal{D}_7}{2}}\dot z^2\ddot z}{N^3} + \frac{\mathcal{D}_4^2}{16}\frac{z^{\frac{\mathcal{D}_9}{2}}\dot z^4}{N^3}  - \frac{\mathcal{D}_4}{2}\frac{z^{\frac{\mathcal{D}_7}{2}}\dot z^3\dot N}{N^4} \bigg)\\
& + 36k^2Nz^{\frac{\mathcal{D}_5}{2}} - 3k\mathcal{D}_1\mathcal{D}_6\frac{z^{\frac{\mathcal{D}_7}{2}}\dot z^2}{N} \bigg\} \bigg]dt + \Sigma_{{R^2}_{D_2}},
\end{split}\ee
where, $\mathcal{D}_i = D-i, ~\text{with},~ i = 1,2\cdots$. At this stage it is customary to introduce a new variable $x = \dot z$, and follow either Ostrogradski's technique by defining momenta as $p_x = {\partial L\over \partial \dot x} ~\mathrm{and}~p_z = {\partial L\over \partial \dot z} - {d p_x\over dt}$, or better to define momenta as usual $p_x = {\partial L\over \partial \dot x} ~\mathrm{and}~p_z = {\partial L\over \partial \dot z}$, resulting in a singular Lagrangian and follow Dirac's constraint analysis. However, in the process, the supplementary boundary term $\Sigma_{{R^2}_{D_2}}$ is not taken care of, and therefore these techniques lack mathematical rigour. Therefore, we introduce the auxiliary variable following Horowitz's prescription \cite{Horowitz} as,
\be\label{Q4} Q ={\partial A_3\over\partial \ddot z}= \frac{\beta\mathcal{D}_1^2 z^{\frac{\mathcal{D}_7}{2}}}{N^3}\left[ 2z\Big(\ddot z - \frac{\dot z\dot N}{N} \Big) + \frac{\mathcal{D}_4}{2}\dot z^2\right]\ee
and express the action (\ref{38}) as,
\be\label{AQz}\begin{split}
A_3 &= \int\bigg[\frac{1}{\kappa}\left(3kNz^{\frac{\mathcal{D}_3}{2}} - \frac{\Delta_1}{8}\frac{z^{\frac{\mathcal{D}_5}{2}}\dot z^2}{N} - \Lambda_0Nz^{\frac{\mathcal{D}_1}{2}} \right) - \gamma\bigg(\frac{\Delta_1\Delta_2}{48}\frac{z^{\frac{\mathcal{D}_9}{2}}\dot z^4}{N^3} + 3k\Delta_2\frac{z^{\frac{\mathcal{D}_7}{2}}\dot z^2}{N} \bigg) \\
&+ Q\ddot z - \frac{{Q}^2N^3}{4\beta\mathcal{D}_1^2z^{\frac{\mathcal{D}_5}{2}}} -\frac{Q\dot z\dot N}{N} + \frac{\mathcal{D}_4}{4}\frac{\dot z^2Q}{z} - 3k\beta\mathcal{D}_1\mathcal{D}_6\frac{z^{\frac{\mathcal{D}_7}{2}}\dot z^2}{N} + 36k^2\beta Nz^{\frac{\mathcal{D}_5}{2}} \bigg]dt + \Sigma_{{R^2}_{D_2}}. \end{split}\ee
\noindent
It is to be mentioned that although $Q$ contains second derivative of $z$, it does not create problem in canonical analysis, since $Q$ is just an auxiliary variable and should be replaced by true canonical variable $K_{ij}$ at the end. Now after removing rest of the total derivative terms under integration by parts once again, which gets cancelled with the remaining boundary term, the action in its final canonical form is expressed as,
\be\label{A20}\begin{split}
A_3 &= \int\bigg[\frac{1}{\kappa}\left(3kNz^{\frac{\mathcal{D}_3}{2}} - \frac{\Delta_1}{8}\frac{z^{\frac{\mathcal{D}_5}{2}}\dot z^2}{N} - \Lambda_0Nz^{\frac{\mathcal{D}_1}{2}} \right) - \gamma\bigg( \frac{\Delta_1\Delta_2}{48}\frac{z^{\frac{\mathcal{D}_9}{2}}\dot z^4}{N^3} + 3k\Delta_2\frac{z^{\frac{\mathcal{D}_7}{2}}\dot z^2}{N} \bigg)  \\
&- \frac{\dot z}{N} (N\dot Q + Q \dot N) - \frac{{Q}^2N^3}{4\beta\mathcal{D}_1^2z^{\frac{\mathcal{D}_5}{2}}} + \frac{\mathcal{D}_4}{4}\frac{\dot z^2Q}{z} - 3k\beta\mathcal{D}_1\mathcal{D}_6\frac{z^{\frac{\mathcal{D}_7}{2}}\dot z^2}{N} + 36k^2\beta Nz^{\frac{\mathcal{D}_5}{2}} \bigg]dt. \end{split} \ee
Canonical momenta are therefore,
\begin{subequations}\begin{align}
p_N &= -\frac{Q \dot z}{N} \label{pq}\\
p_Q &= - \dot z, \label{pq1}\\
p_z &= - \frac{\Delta_1}{4\kappa}\frac{z^{\frac{\mathcal{D}_5}{2}}\dot z}{N} - \dot Q -\frac{Q\dot N}{N} + \frac{\mathcal{D}_4}{2}\frac{\dot zQ}{z} - 6k\beta\mathcal{D}_1\mathcal{D}_6\frac{z^{\frac{\mathcal{D}_7}{2}}\dot z}{N} - \gamma\bigg(\frac{\Delta_1\Delta_2}{12}\frac{z^{\frac{\mathcal{D}_9}{2}}\dot z^3}{N^3} + 6k\Delta_2\frac{z^{\frac{\mathcal{D}_7}{2}}\dot z}{N} \bigg). \label{pq2}
\end{align}\end{subequations}
Under ${(D-1)+1}$ decomposition, the in-built general covariance of the theory of gravity leads to the so called diffeomorphic invariance. This means that the lapse function $N$ should act as Lagrange multiplier and variation with respect to $N$ would give Hamiltonian constraint $(H_c = 0)$. Therefore, here we present only the $N$ variation equation viz.,

\be\label{N}\begin{split}
& -\frac{Q\ddot z}{N} - \frac{\dot Q\dot z}{N} + \frac{3{Q}^2N^2}{4\beta\mathcal{D}_1^2z^{\frac{\mathcal{D}_5}{2}}} - 36k^2\beta z^{\frac{\mathcal{D}_5}{2}} - 3k\beta\mathcal{D}_1\mathcal{D}_6\frac{z^{\frac{\mathcal{D}_7}{2}}\dot z^2}{N^2} -\frac{1}{\kappa}\left(3kz^{\frac{\mathcal{D}_3}{2}} + \frac{\Delta_1}{8}\frac{z^{\frac{\mathcal{D}_5}{2}}\dot z^2}{N^2} - \Lambda_0z^{\frac{\mathcal{D}_1}{2}} \right) \\
&- \gamma\bigg(\frac{\Delta_1\Delta_2}{16}\frac{z^{\frac{\mathcal{D}_9}{2}}\dot z^4}{N^4} + 3k\Delta_2\frac{z^{\frac{\mathcal{D}_7}{2}}\dot z^2}{N^2} \bigg)= 0.
\end{split}\ee
Note that unlike general theory of Relativity, the action contains $\dot N$ and so $p_N \ne 0$. Nevertheless, the Lagrangian is singular as the determinant of Hessian vanishes, signalling the presence of a constraint $(H_c = 0)$. If one fixes $N = 1$ say, from the beginning, Lagrangian is no longer degenerate and one loses Hamilton constraint equation. In that case, one has to find the Hamiltonian and set it equal to zero. However, the appearance of $\dot N$ term in the action is due to bad choice of auxiliary variable, since under a different choice, viz. $q = N Q$, $N\dot Q  + \dot N Q = {\dot q}$, and so $\dot N$ disappears from the action (\ref{A20}). Nevertheless, even without introducing $q$, one can handle the situation by eliminating $\ddot z$ between definition of $Q$ presented in equation (\ref{Q4}) and the $N$ variation equation (\ref{N}). In the process, the Hamiltonian of the system reads
\be\label{Hc5iso1}\begin{split}
H & =  - \dot Q\dot z - \frac{Q\dot z\dot N}{N} + \frac{Q^2N^3}{4\beta\mathcal{D}_1^2z^{\frac{\mathcal{D}_5}{2}}} - \frac{\mathcal{D}_4}{4}\frac{\dot z^2Q}{z}  - 36k^2N\beta z^{\frac{\mathcal{D}_5}{2}} - 3k\beta\mathcal{D}_1\mathcal{D}_6\frac{z^{\frac{\mathcal{D}_7}{2}}\dot z^2}{N}\\
&-\frac{N}{\kappa}\left(3kz^{\frac{\mathcal{D}_3}{2}} + \frac{\Delta_1}{8}\frac{z^{\frac{\mathcal{D}_5}{2}}\dot z^2}{N^2} - \Lambda_0z^{\frac{\mathcal{D}_1}{2}} \right) - \gamma\bigg(\frac{\Delta_1\Delta_2}{16}\frac{z^{\frac{\mathcal{D}_9}{2}}\dot z^4}{N^3} + 3k\Delta_2\frac{z^{\frac{\mathcal{D}_7}{2}}\dot z^2}{N} \bigg),
\end{split}\ee
which is constrained to vanish. Now, in view of the definitions of momenta (\ref{pq1}) and (\ref{pq2}), it is possible to construct the phase-space formulation of the Hamiltonian constraint equation as,
\be\label{Hc5iso2}\begin{split}
H & =  -p_Qp_z + \frac{Q^2N^3}{4\beta\mathcal{D}_1^2z^{\frac{\mathcal{D}_5}{2}}} - \frac{\mathcal{D}_4}{4}\frac{Q p_Q^2}{z} - 36k^2N\beta z^{\frac{\mathcal{D}_5}{2}} + 3k\beta\mathcal{D}_1\mathcal{D}_6\frac{z^{\frac{\mathcal{D}_7}{2}}p_Q^2}{N}  \\
&-\frac{N}{\kappa}\left(3kz^{\frac{\mathcal{D}_3}{2}} - \frac{\Delta_1}{8}\frac{z^{\frac{\mathcal{D}_5}{2}}p_Q^2}{N^2} - \Lambda_0z^{\frac{\mathcal{D}_1}{2}} \right) + \gamma\bigg(\frac{\Delta_1\Delta_2}{48}\frac{z^{\frac{\mathcal{D}_9}{2}}p_Q^4}{N^3} + 3k\Delta_2\frac{z^{\frac{\mathcal{D}_7}{2}}p_Q^2}{N} \bigg)= 0
\end{split}\ee
The definition of momentum $p_z$ presented in (\ref{pq2}) still indicates that it is multivalued in $\dot z$, but $p_Q$ given in (\ref{pq1}) being a single valued function of $\dot z$, $p_z$ turns out to be single valued in $\dot Q$. Thus standard Legendre transformation is admissible. In the process, the presence of higher order term alleviates the problem associated with branched Hamiltonian, presenting a unique Hamiltonian (\ref{Hc5iso2}) in phase space, quite naturally. This is a common feature of curvature squared gravity theory and here such an additional $R^2$ term cures the disease of the lack in Hamiltonian structure of the Lanczos-Lovelock action.

\subsection{Diffeomorphic invariance of the action}

As mentioned in the introduction, diffeomorphic invariance ($H = N{\mathcal H}$) is not manifest in Lanczos-Lovelock gravity. However, the Hamltonian (\ref{Hc5iso2}) admits diffeomorphic invariance, in a straight forward manner, once it is expressed in terms of the basic variables $(h_{ij}, K_{ij})$. For this purpose, we choose
\be K_{ij} \equiv x = \frac{\dot z}{N} \ee
and replace $Q$ and $p_Q$ by,
\begin{subequations}\begin{align}
Q &= \frac{\partial A}{\partial \ddot z} = \frac{p_x}{N}\label{rel} \\
p_{Q} &= -\dot z = -N x
\end{align}\end{subequations}
following Horowitz \cite{Horowitz}. The above transformations from the phase space variables ($Q, p_Q$) to ($\dot z, p_{\dot z}$) or ($x, p_x$) in particular, are canonical. Therefore, the Hamiltonian constraint equation (\ref{Hc5iso2}) now takes the form,
\be\label{Hc5iso3}\begin{split}
H &= N\bigg[xp_z + \frac{p_x^2}{4\beta\mathcal{D}_1^2z^{\frac{\mathcal{D}_5}{2}}} - \frac{\mathcal{D}_4x^2p_x}{4z} - 36k^2\beta z^{\frac{\mathcal{D}_5}{2}} + 3k\beta\mathcal{D}_1\mathcal{D}_6z^{\frac{\mathcal{D}_7}{2}}x^2  \\
&+ \frac{1}{\kappa}\left(\frac{\Delta_1}{8}z^{\frac{\mathcal{D}_5}{2}}x^2 - 3kz^{\frac{\mathcal{D}_3}{2}} + \Lambda_0z^{\frac{\mathcal{D}_1}{2}} \right) + \gamma \bigg(\frac{\Delta_1\Delta_2}{48}z^{\frac{\mathcal{D}_9}{2}}x^4 + 3k\Delta_2z^{\frac{\mathcal{D}_7}{2}}x^2 \bigg) \bigg] = N{\mathcal{H}} = 0.
\end{split}\ee
The action (\ref{AQz}) can now be expressed in the ADM form \cite{ADM} with respect to the basic variables as,

\be\begin{split} A_3 &= \int\left(\dot z p_z + \dot x p_x - N\mathcal{H}\right)dt~ d^3 x = \int\left(\dot h_{ij} \pi^{ij} + \dot K_{ij}\Pi^{ij} - N\mathcal{H}\right)dt~ d^3 x,\end{split}\ee
where, $\pi^{ij}$ and $\Pi^{ij}$ are momenta canonically conjugate to $h_{ij}$ and $K_{ij}$ respectively. Hence, diffeomorphic invariance of the action is now manifest. Thus we observe that incorporating higher order ($R^2$) term, the modified Lanczos-Lovelock gravity (\ref{A4}) becomes free from all pathological behaviour. It is not difficult to show that in the limit $\beta = 0$, the pathological behaviour of Lanczos-Lovelock gravity reappears and further, in the limit $\gamma = 0$, General Theory of Relativity is reproduced (see Appendix A).

\subsection{Canonical Quantization}

Canonical quantization of the above Hamiltonian constraint equation (\ref{Hc5iso3}) reads
\be\label{qeq}\begin{split}
& i\hbar z^{\frac{D_5}{2}}\frac{\partial \Psi}{\partial z} = -\frac{\hbar^2}{4 \beta \mathcal{D}_1^2}\frac{1}{x}\left(\frac{\partial^2}{\partial x^2} + \frac{n}{x}\frac{\partial}{\partial x}\right)\Psi + \frac{i\hbar}{8} \mathcal{D}_4 z^{\frac{D_7}{2}}\left(x\frac{\partial \Psi}{\partial x}+{\partial\over \partial x}(\Psi x) \right)\\
& + \Bigg[ 3k\beta\mathcal{D}_1\mathcal{D}_6z^{\mathcal{D}_6}x - 36k^2\beta \frac{z^{\mathcal{D}_5}}{x} +\gamma \Bigg( \frac{\Delta_1\Delta_2}{48}z^{\mathcal{D}_7}x^3 + 3k\Delta_2z^{\mathcal{D}_6}x \Bigg) + {1 \over \kappa}\Bigg(\frac{\Delta_1}{8}z^{\mathcal{D}_5}x - 3k\frac{z^{\mathcal{D}_4}}{x} + \Lambda_0\frac{z^{\mathcal{D}_3}}{x} \Bigg)\Bigg]\Psi,
\end{split}\ee
where, $n$ is the operator ordering index, and in the second term we have used Weyl ordering. Under a further change of variable $\big(\alpha = z^{-\frac{\mathcal{D}_7}{2}}\big)$, equation (\ref{qeq}) takes the appearance of the Schr\"odinger equation
\be\begin{split}\label{qef}
i\hbar\frac{\partial \Psi}{\partial \alpha}& = \frac{\hbar^2}{2\beta \mathcal{D}_1^2\mathcal{D}_7 x}\Big(\frac{\partial^2}{\partial x^2} + \frac{n}{x}\frac{\partial}{\partial x}\Big)\Psi - \frac{i\hbar}{4} \frac{\mathcal{D}_4}{\mathcal{D}_7}\frac{1}{\alpha}\left(2x\frac{\partial \Psi}{\partial x}+\Psi \right)+ V_e(x, \alpha)\Psi = \hat H_e(x,\alpha)\Psi.
\end{split}\ee
Here $\hat H_e$ is the effective Hamiltonian and $``\alpha"$ plays the role of internal time parameter. The effective potential $V_e$ is given by,
\be\label{Ve}\begin{split}
V_e(x, \alpha) &= {1\over \mathcal{D}_7}\Bigg[\frac{72k^2\beta}{x \alpha^{\frac{2\mathcal{D}_5}{\mathcal{D}_7}}} - 6k\beta\frac{\mathcal{D}_1\mathcal{D}_6x}{\alpha^{\frac{2\mathcal{D}_6}{\mathcal{D}_7}}} - \gamma \left( \frac{\Delta_1\Delta_2}{24}\frac{x^3}{\alpha^2} + 6k\Delta_2\frac{x}{\alpha^{\frac{2\mathcal{D}_6}{\mathcal{D}_7}}} \right) - {1 \over \kappa x}\left(\frac{\Delta_1}{4}\frac{x^2}{\alpha^{\frac{2\mathcal{D}_5}{\mathcal{D}_7}}} - \frac{6k}{ \alpha^{\frac{2\mathcal{D}_4}{\mathcal{D}_7}}} + \frac{2\Lambda_0}{ \alpha^{\frac{2\mathcal{D}_3}{\mathcal{D}_7}}} \right)\Bigg] .
\end{split}\ee
The very first point to note is that, the contribution of Gauss-Bonnet term appears only in the potential. The second important feature is, in dimension $D=4$, $\mathcal{D}_4 (= D-4)$ vanishes, and the main contribution appearing from dimensions higher than $4$, viz., the second term on the right hand side of equation (\ref{qef}) vanishes. Further, $\Delta_2 = \mathcal{D}_3\mathcal{D}_4$ also vanishes and hence Gauss-Bonnet term does not contribute. As a result, the Hamiltonian operator (\ref{qef}) takes exactly the form obtained earlier in \cite{aks4}. More interesting feature is, for $\mathcal{D} < 7$, as the so called time parameter ``$\alpha$" increases, with the expansion of the universe, contribution from the second term of (\ref{qef}) becomes negligible, leading to natural compactification to four dimension. However, for $D = 7$, the potential term blows, while for $D > 7$, the time parameter increases as the scale-factor decreases, indicating a contracting model. Unfortunately, contribution from higher dimension vanishes once again, as the universe collapses, which is un-physical. Therefore, for some reason (unknown to the present authors) the Schr\"odinger-like equation (\ref{qef}) holds for $D < 7$. Nevertheless, equation (\ref{qeq}) is free from all such trouble.\\

It should be mentioned that the Schr\"odinger-like equation could have also been obtained simply dividing (\ref{qeq}) by $z^{D_5\over 2}$. However, the motivation behind all the manipulations made in equation (\ref{qeq}) was to keep the very first term in (\ref{qef}) free from $\alpha$. As a result, due to the time dependence of the effective Hamiltonian $\hat H_e(x,\alpha)$, the above Schr\"odinger-like equation (\ref{qef}) may now be treated in the interaction picture as,
\begin{subequations}\begin{align}
\hat H_e &= \hat H_0 + \hat H_I, \ \ \ \text{with} \label{int}\\
\hat H_0 &= \frac{\hbar^2}{2\beta \mathcal{D}_1^2\mathcal{D}_7 x}\Big(\frac{\partial^2}{\partial x^2} + \frac{n}{x}\frac{\partial}{\partial x}\Big) \\
\hat H_I &= - \frac{i\hbar}{4} \frac{\mathcal{D}_4}{\mathcal{D}_7}\frac{1}{\alpha}\left(2x\frac{\partial}{\partial x}+1\right)+ V_e(x, \alpha)
\end{align}\end{subequations}
where, $\hat H_I$ is the interacting term. $\hat H_0$ is hermitian for a particular choice of operator ordering index $n = -1$, and the first term in $\hat H_I$, has been made hermitian under Weyl ordering (see appendix B). So, clearly $\hat H_e$ is hermitian. Since, $\hat H_0$ is now solvable, the eigen-decomposition are known, and so one can find the propagator as $U = e^{-{i\over \hbar}\hat H_e\alpha}$. Now, as $\hat H_I$ is also hermitian, one may be tempted to write $U_I = e^{-{i\over \hbar}\int _0^t \hat H_I(\alpha') d\alpha'}$. However, note that although both the terms in $\hat H_I$ individually commute at two different epoch ($\alpha_1, \alpha_2$), $\hat H_I$ as a whole does not, i.e. $[\hat H_I(\alpha_1), \hat H_I(\alpha_2)] \ne 0$. Therefore, we don't get explicit solution in terms of an integral. Nevertheless, we can find formal solution in Dyson interacting picture, so that the propagator takes the form
\be U(\alpha,\alpha_0) = \mathcal{T}e^{-{i\over \hbar}\int_{\alpha_0}^{\alpha}\hat H_I d\alpha}\ee
$\mathcal{T}$ being the time ordering operator. This is the expression of unitary operator for the time-dependent Hamiltonian under consideration, and the unitarity of extended Lanczos-Lovelock gravity has been established.\\

\noindent
The hermiticity of $\hat H_e$ now allows one to write the continuity equation , as,
\be \frac{\partial\rho}{\partial \alpha} + \nabla . {\bf{J}} = 0, \ee
where, $ \rho = \Psi^*\Psi ~~ \text{and} ~~  {\bf J} = ({\bf J}_x, 0, 0) $ are the probability density and the current density respectively, with,
\be {\bf J}_x = \frac{i \hbar }{2 \beta \mathcal{D}_1^2 \mathcal{D}_7}{1\over x}(\Psi^*\Psi_{,x}-\Psi^*_{~,x}\Psi) + {\mathcal{D}_4\over 2\mathcal{D}_7}{x\over \alpha}\Psi^*\Psi. \ee
In the process, operator ordering index here too has been fixed as $n = -1$ from physical argument. Note that under the choice $\gamma = 0$, the Gauss-Bonnet term disappears and the results obtained earlier in \cite{aks5} are recovered, provided instead of $z = {h_{ij}}^{D\over 4}$, one would have started with $z = h_{ij}$.

\subsection{Classical solution and Semiclassical approximation}

Now a viable quantum theory should reproduce the original classical scenario in appropriate limits under certain semiclassical approximation. Thus, it is now left to be shown that the present quantum prescription admits a viable semiclassical approximation. For this purpose we first find a classical solution to the field equations under consideration. Clearly, either form of the Hamiltonian constraint equation (\ref{Hc5iso1}), (\ref{Hc5iso2}) or (\ref{Hc5iso3}) admits de-Sitter solution (for $k=0$) in the form
\be \label{clsol} a = a_0\exp{( {\mathrm H} t)}, \ee
provided,
\be\begin{split}
& \mathcal{D}_2 + 2\kappa {\mathrm H}^2\mathcal{D}_4\Bigg[4D\mathcal{D}_1\beta + \mathcal{D}_2\mathcal{D}_3\gamma\Bigg] = \frac{2\Lambda_0}{\mathcal{D}_1{\mathrm H^2}},
\end{split}\ee
where, ${\mathrm H}$ is a constant. Next, to present semiclassical solution in the standard WKB approximation, let us, for the sake of simplicity, take up the time-independent equation (\ref{qeq}) and express it as
\be\begin{split}\label{semid} &-\frac{\hbar^2}{4 \beta \mathcal{D}_1^2}\frac{1}{z^{\frac{D_5}{2}}}\left(\frac{\partial^2}{\partial x^2} + \frac{n}{x}\frac{\partial}{\partial x}\right)\Psi  - i\hbar x\frac{\partial \Psi}{\partial z} + i\hbar \frac{\mathcal{D}_4x^2}{4z}\frac{\partial \Psi}{\partial x} + V\psi = 0,\end{split}\ee
where
\be \begin{split}
V &= - 36k^2\beta z^{\frac{\mathcal{D}_5}{2}} + 3k\beta\mathcal{D}_1\mathcal{D}_6z^{\frac{\mathcal{D}_7}{2}}x^2 + \frac{1}{\kappa}\left(\frac{\Delta_1}{8}z^{\frac{\mathcal{D}_5}{2}}x^2 - 3kz^{\frac{\mathcal{D}_3}{2}} + \Lambda_0z^{\frac{\mathcal{D}_1}{2}} \right) + \gamma \bigg(\frac{\Delta_1\Delta_2}{48}z^{\frac{\mathcal{D}_9}{2}}x^4 + 3k\Delta_2z^{\frac{\mathcal{D}_7}{2}}x^2 \bigg)
\end{split}\ee
The above equation may be treated as time independent Schr{\"o}dinger equation with two variables $x$ and $z$ and therefore, as usual, let us sought the solution of equation (\ref{semid}) as,
\be\label{psisemi} \psi = \psi_0e^{\frac{i}{\hbar}S(x,z)}\ee

\noindent
and expand $S$ in power series of $\hbar$ as,
\be\label{S0} S = S_0(x,z) + \hbar S_1(x,z) + \hbar^2S_2(x,z) + .... \ .\ee
Now following our earlier work \cite{aks4} the semiclassical wavefunction around the classical solution reads (up to first order approximation)
\be \psi =A_0 e^{\frac{i}{\hbar}\left[-\frac{\Delta_1{\mathrm H}}{\kappa\mathcal{D}_1} - 4\beta D\mathcal{D}_1{\mathrm H}^3 - \frac{4}{3}\gamma \Delta_1\Delta_2H \right]z^{\frac{\mathcal{D}_1}{2}}} \ee
where, $A_0 = \psi_0 \left\{\left(\beta D{\mathcal{D}_1}^2 \right)^{-\frac{D}{16}} z^{-\frac{\mathcal{D}_5(D^2-16D+256)}{32D}}\right\}$. The wavefunction is clearly oscillatory being peaked around the classical solution (\ref{clsol}). Thus we have administered all the fundamental features of a viable quantized theory corresponding to the modified Lanczos-Lovelock action.

\section{Spontaneous breaking of time translational symmetry and its remedy}

As demonstrated by Shapere and Wilczek \cite{shap}, when a physical solution of a set of equations displays less symmetry than the equations itself, then the solution is said have broken the symmetry spontaneously. Now, every classical conservative system is associated with a conserved total energy $(H)$, which generates a continuous time translational symmetry. However, minimizing the total energy, if it is observed that the ground state (minimum energy state) is associated with non-trivial motion of the ground state coordinates, then the symmetry is spontaneously broken. For example, in the case of Harmonic oscillator, $H = {p_x^2\over 2m} + kx^2$ is conserved. Nevertheless, one can still minimize it to $H_{min} = 0$, but this minima is located at $\dot x = {\partial H\over \partial p_x} =0$ and $\dot p_x = - {\partial H\over \partial x} =0$, which correspond to no motion at all and so the symmetry is restored. To demonstrate how time translational symmetry (TTS) is broken spontaneously in the case of branched Hamiltonian and can't be restored by any technique of canonical formulation adopted so far, let us first take the toy model (\ref{chL}), for which the corresponding energy function is
\be E = {3\over 4}\dot \phi^2 - {1\over 2}\dot \phi^2\ee
Now minimization of the energy function yields
\be {\partial E\over \partial \dot \phi} = {(3\dot\phi^2 - 1)\dot \phi}.\ee
One can check that $\dot \phi = 0$ leads to a maximum, while, $(3\dot\phi^2 - k)$ leads to a minimum. However, at the minimum, $E_{min} = -{1\over 12}$, and the ground state coordinate is $\dot \phi^G = \pm\sqrt {1\over 3}$, yielding $\phi^G = \pm\sqrt{1\over 3}t  + \phi_0$, which dictates non-trivial motion of the ground state. Hence, TTS is spontaneously broken.\\

\noindent
He and Chi \cite{Chi} have demonstrated in view of figure-1, that the ground state Hamiltonian obtained following Legendre-Fenchel transformation is $H_0 = {1\over 4}$, being located at $p^G = 0$ and $\dot\phi^G = \pm 1$. Hence TTS is broken spontaneously, even after constructing the single valued Hamiltonian. In the technique adopted by Avraham and Bustein \cite{avra} corresponding to the same toy model (\ref{chL}), the Hamiltonian is $H = {3\over 4} g^4 - {1\over 2} g^2$, where, $g = \dot \phi$ has been treated as non-Darboux momentum. Here again, the minima of the Hamiltonian, $H_{min} =  -{1\over 12}$ is located at $g^G = \dot \phi^G = \pm\sqrt{1\over 3}$ and the TTS is again spontaneously broken. As already mentioned, following a slightly different route, Zhao, Yu and Xu \cite{3.9} have obtained the same Hamiltonian as Avraham and Bustein \cite{avra}, and so their method also lead to the same result. A physical interpretation of such uncanny behaviour has been given recently \cite{shap}. As the spatial periodicity is associated with the formation of ordinary crystals, likewise, Shapere and Wilczek \cite{shap} referred it to the formation of time crystals. However, such interpretation is debatable, since it doesn't hold for gravity.\\

\noindent
Before turning our attention to the more realistic situation, i.e. Lanczos-Lovelock gravity, let us examine how such uncanny situation may be remedied under the introduction of a higher order term. For this purpose, we express the modified Lagrangian as
\be L = {1\over 4}\dot \phi^4 - {1\over 2}\dot \phi^2 + \alpha\ddot\phi^2.\ee
One can now introduce an auxiliary variable $q = {\partial L\over\partial \ddot\phi} = 2\alpha\ddot\phi$, to obtain the point Lagrangian as,
\be L = {1\over 4}\dot \phi^4 - {1\over 2}\dot \phi^2 + \dot q\dot\phi - {q^2\over 4\alpha}.\ee
The energy function is then expressed as
\be E = {3\over 4}\dot \phi^4 - {1\over 2}\dot \phi^2 + \dot q\dot\phi + {q^2\over 4\alpha}.\ee
It is now possible to check that the ground state energy is $E_{min} = -{1\over 12}$, and is located at $\dot q^G = 0 = q^G$ and $\dot \phi^G = \pm \sqrt{1\over 3}$. Hence, TTS is spontaneously broken. Now the Hamiltonian of the system is
\be H = p_qp_{\phi} - {1\over 4}p_q^4 +{1\over 2}p_q^2 + {q^2\over 4\alpha}\ee
which, under the choice $\dot\phi = x$ may be transformed in terms of the basic variables ($\phi, p_{\phi}; x, p_x$) under the replacement $q \rightarrow p_x$ and $p_q \rightarrow x$ as
\be H = xp_{\phi} + {p_x^2\over 4\alpha} -{1\over 4}x^4 + {1\over 2} x^2.\ee
The Hamilton's equations of motion are
\be \dot x = {\partial H\over \partial p_x} = {p_x\over 2\alpha}; \dot p_x = -{\partial H\over \partial x} = -p_{\phi} + x(x^2 - 1);\;\;\dot\phi = {\partial H\over \partial p_{\phi}} = x;  \dot p_{\phi} = -{\partial H\over \partial \phi} = 0.\ee
One can easily check that the minima,  $H_{min} = 0$ is now located at $x^G = 0 = p_x^G$, which correspond to $\dot\phi^G = 0, \dot x^G = 0, \dot p_x = 0, \dot p_{\phi} = 0$. Therefore, TTS is restored.\\

\noindent
Let us now turn our attention to the more realistic situation. For the purpose of demonstration, we work with action (\ref{AA4}) and further choose $N = 1, \Lambda_0 = 0 = k$, to avoid complication. The point Lagrangian therefore is
\be L = - {6\over \kappa}a^2\dot a^2 - 8\gamma\dot a^4.\ee
The canonical momenta is
\be p_a = -{12a^2\dot a\over \kappa} - 32 \gamma\dot a^3\ee
and the energy function is
\be E = -24\gamma \dot a^4 - {6\over \kappa}a^2\dot a^2.\ee
It is easy to check that the minima, $E_{min} = {3 a^2 \over 8\gamma \kappa^2}$ is located at $a^G = a_0\exp\left({\pm {i\over\sqrt {8\gamma\kappa}}}\right)t$. Now, just the same above interpretation that `the ground state has non-trivial motion and TTS is spontaneously broken giving rise to the formation of time crystal in the classical domain \cite{shap}' is not enough here. The situation is even disastrous. The Hamiltonian (the total energy) in gravity is constrained to vanish due to diffemorphism. So, it does not admit deviation from this, under any circumstances. If the energy is set to vanish, then $a = 0$, leading to contradiction. Thus, not that diffeomorphic invariance is not manifest, rather Lanczos-Lovelock gravity (\ref{A3}) lacks diffeomorphic invariance, which is an incurable disease. However, diffeomorphic invariance is manifest in the action (\ref{A4}), since otherwise classical field equations are not satisfied, which has been depicted in the modified ADM action (\ref{AQz}). As diffeomorphic invariance constraints Hamiltonian to vanish, question of spontaneous breaking of time translational symmetry doesn't arise at all. However, such constraint only restricts the ground state to inflationary solution in the form
\be a = a_0 e^{\sqrt{3\over 2\beta}t}\ee
which is the most attractive feature of $R^2$ term, being dubbed as curvature induced inflation.

\section{Summary}
If a Lagrangian contains velocity higher than quadratic, inversion of velocity in terms of momentum leads to multivalued Hamiltonian with cusps at minima. If the velocity is non-vanishing at the cusp, the lowest energy solution of such system spontaneously breaks time translation. Lanczos-Lovelock gravity suffers from such disease of having branched Hamiltonian. Several techniques appear in the literature, to get rid of the difficulty. Every technique has its limitation, and even the phase-space Hamiltonian doesn't always produce correct Euler-Lagrange equations in some situations. Further, it is important to mention that no two Hamiltonians for the same system obtained following different techniques, are related through canonical transformation. The final test to check if a Hamiltonian is correct, is to quantize the system and to see if under some appropriate semiclassical approximation, the behaviour of the wave-function is oscillatory about a known classical solution. This has not been performed in the literature and as we observe, is impossible with the Hamiltonian presented by different authors. Thus, the issue of branched Hamiltonian is far from being resolved.\\

All the attempts to cure the pathology of branched Hamiltonian associated with Lanczos-Lovelock gravity has been made in the spatially flat ($k = 0$) Robertson-Walker minisuperspace model, and no technique could restore time-translational symmetry. This is due to the fact that the theory lacks diffeomorphic invariance. To get round the difficulty, we propose that if higher degree terms appear with higher order in the action, then appropriate canonical formulation of higher order theory removes the pathology of branched Hamiltonian. In the process, time translation symmetry remains preserved. Therefore, we have modified the Lanczos-Lovelock action adding a scalar curvature squared term. Unitarity of Lanczos-Lovelock gravity has been established perturbatively. This has been done by finding its propagators which arise from $h^2$ expansion of the theory around a fixed Minkowskian (flat) background $g_{\mu\nu} =  \eta_{\mu\nu} + h_{\mu\nu}$ ($\eta_{\mu\nu}$ is the Minkowski metric, $h_{\mu\nu}$ being the perturbation with $h = h^{\mu}_{\mu}$), and also around other fixed backgrounds. Nevertheless, its fate was not known non-perturbatively in the context of quantum cosmology (say), since it lacks a Hamiltonian structure in canonical form. Modified Lanczos-Lovelock gravity regulates the issue of branching and produces an unique Hamiltonian structure accompanying a unitary evolution operator. Note that a renormalized theory of gravitation in $4$-dimension requires curvature squared terms in the action. It is $R_{\mu\nu}R^{\mu\nu}$ term, which is responsible for ghosts and not $R^2$, as it only gives a massive scalar mode. In this connection one may also note that, in homogeneous and isotropic space-time, the most general form of an action viz.,
\be\begin{split} A &= \int \sqrt{-g} d^D x \left[{R-2\Lambda_0\over 2\kappa} + \beta_1 R^2 + \beta_2I_1 + \beta_3 I_2\right]\end{split}\ee
reduces to
\be A = \int \sqrt{-g} d^D x \left[{R-2\Lambda_0\over 2\kappa} + \beta R^2 + \gamma {\mathcal G}_D\right]\ee
since, $R_{\mu\nu}R^{\mu\nu} - {D\over 4(D-1)}R^2$ is a total derivative term in the isotropic and homogeneous space-time (where $I_1 = R_{\mu\nu}R^{\mu\nu}$ and $I_2 = R_{\mu\nu\delta\sigma}R^{\mu\nu\delta\sigma}$). Unitarity of the above action has been established in the fixed curved dS/AdS background, under the constraint
\be\frac{1}{\kappa_e} = \frac{1}{2\kappa} + \frac{4\Lambda D}{D-2}\beta +\frac{4\Lambda\Delta_2}{\Delta_1}\gamma \ee
for $D \ge 3$ \cite{unitary3} . In the above, $\Lambda$ is the effective cosmological constant which is related to the bare cosmological constant$\Lambda_0$, by the following relation,
\be\frac{\Lambda-\Lambda_0}{4\kappa} + \left[\frac{D(D-4)}{(D-2)^2}\beta + \frac{\Delta_2}{\Delta_1}\gamma\right]\Lambda^2 = 0,\ee
while, the mass of the scalar mode in dS and AdS backgrounds should satisfy following conditions respectively,
\begin{subequations}\begin{align}
& m_s^2 = \frac{D-2}{4(D-1)\beta\kappa_e} - 2\frac{\Lambda D}{\Delta_1}\ge 0 \\
& m_s^2 \ge \frac{D-1}{\Delta_1\Lambda},~~ \text{with} ~~\kappa_e > 0.
\end{align}\end{subequations}
Therefore, despite the usual thought that (due to the absence of fourth derivative terms in the field equations) if the propagator of a higher order theory reduces to that corresponding to Lanczos-Lovelock gravity in generic $D$-dimension, then only the theory is unitary, the analysis in \cite{unitary3} clearly admits $R^2$ term in addition. Hence, the extended Lanczos-Lovelock action is free from all pathologies.\\

\noindent
So far, all the attempts to resolve the issue of branched Hamiltonian have been made with finite degrees of freedom. It might therefore appear that the resolution of the  problem here, has been possible only by suppressing large number of degrees of freedom associated with the whole superspace. However, as already mentioned in the introduction, Hamiltonian formulation of $f(Riemann) \propto C_{\alpha\beta\mu\nu}C^{\alpha\beta\mu\nu}$ theories of gravity ($C_{\alpha\beta\mu\nu}$ being the Weyl tensor, which contains Gauss-Bonnet term (4) along with curvature squared terms), has recently been performed in the whole superspace \cite{Deruelle} which does not show any pathology. This clearly dictates that resolution of the pathology of branched Hamiltonian together with the spontaneously broken time translational symmetry is generic and not restricted to finite degrees of freedom. We have considered minisuperspace, to express the modified Wheeler-deWitt equation in the form of Schr\"odinger equation, which results in quantum mechanical probabilistic interpretation and establishes unitary evolution of quantum states. Further, the example cited with point Lagrangian (65) in section V, clearly demonstrates that the presence of higher order term resolves the issue. Thus any higher order curvature invariant term can resolve the issue of branched Hamiltonian. However, one has to be careful that the higher order term introduced for the purpose doesn't suffer from its own pathology. For example, $R_{\mu\nu}^2$ term leads to ghost degrees of freedom when expanded about flat Minkowsi background. Therefore, it is always safe to handle the situation with scalar curvature squared term. Thus, we conclude that higher order theory generically cures the problem associated with branched Hamiltonian. Finally, one might ask, `` can the present study be extended to more general $F(G)$ gravities, where $G$ is the Gauss-Bonnet scalar"? This may be possible for some specific value of $n$, under the choice $F(G) \propto G^n$. However, in general, i.e. for arbitrary $n$, it is not possible. This is because, canonical formulation of $F(G)$ gravity is only possible, following Lagrange multiplier technique, treating $G$ as an auxiliary variable. Canonical transformation from the set of variables ($G, p^G$) to the set of basic variables ($K_{ij}, \Pi^{ij}$) doesn't exist.

\appendix
\section{The GR limit of the Hamiltonian (\ref{Hc5iso3})}
It is important to note that $\beta = 0$ can not be substituted directly in the Hamiltonian (\ref{Hc5iso3}) as it appears in the denominator too. This is because, to handle fourth order gravity one requires to choose a non-zero auxiliary variable $Q \ne 0$. In the absence of $R^2$ term, auxiliary variable is redundant, which is also apparent from the definition (\ref{Q4}), since, $Q = 0$, as $\beta = 0$. Now the Hamiltonian (\ref{Hc5iso3}) contains momenta $p_x$ and $p_z$. The first one vanishes in view of its definition (\ref{rel}), while $p_z$ given in (\ref{pq2}) changes appreciably to,
\begin{equation}\label{Apz}
p_z = - \frac{\Delta_1}{4\kappa}\frac{z^{\frac{\mathcal{D}_5}{2}}\dot z}{N} - \gamma \Delta_2\bigg(\frac{\Delta_1}{12}\frac{z^{\frac{\mathcal{D}_9}{2}}\dot z^3}{N^3} + 6k\frac{z^{\frac{\mathcal{D}_7}{2}}\dot z}{N} \bigg)
\end{equation}
So the Hamiltonian constraint equation (\ref{Hc5iso3}) now reads
\be\begin{split}
\label{AH}H_c &= N\bigg[xp_z + \frac{1}{\kappa}\left(\frac{\Delta_1}{8}z^{\frac{\mathcal{D}_5}{2}}x^2 - 3kz^{\frac{\mathcal{D}_3}{2}} + \Lambda_0z^{\frac{\mathcal{D}_1}{2}} \right) + \gamma \bigg(\frac{\Delta_1\Delta_2}{48}z^{\frac{\mathcal{D}_9}{2}}x^4 + 3k\Delta_2z^{\frac{\mathcal{D}_7}{2}}x^2 \bigg) \bigg] \\
&= \frac{1}{\kappa}\left(-\frac{\Delta_1}{8}\frac{z^{\frac{\mathcal{D}_5}{2}}\dot z^2}{N^2} - 3kz^{\frac{\mathcal{D}_3}{2}} + \Lambda_0z^{\frac{\mathcal{D}_1}{2}} \right) - \gamma \Delta_2\bigg(\frac{\Delta_1}{16}\frac{z^{\frac{\mathcal{D}_9}{2}}\dot z^4}{N^4} + 3k\frac{z^{\frac{\mathcal{D}_7}{2}}\dot z^2}{N^2} \bigg)
\end{split}\ee
Here, we have substituted $x=\frac{\dot z}{N}$ in view of the definition of $x$ given in equation (70) and the expression of $p_z$ given in equation (\ref{Apz}). Now if we further replace $z$ by $a^2$, the Hamiltonian (\ref{AH}) reduces to the one presented in (7), which suffers from the pathological issue of branched Hamiltonian. Clearly, setting $\gamma=0$, GR limit is obtained, and the corresponding Hamiltonian reads
\be
H_c = \frac{N}{\kappa}\Bigg[-{\kappa^2p_a^2 \over 2\Delta_1a^{\mathcal{D}_3}} - 3k a^{\mathcal{D}_3} + \Lambda_0 a^{\mathcal{D}_1}\Bigg] = 0.\ee

\section{To show that the Hamiltonian $\hat H_e$ (\ref{qef}) is hermitian}
First let us take the Hamiltonian $\hat H_0$ appearing in (77b), where $\widehat{T} = \frac{1}{x}\frac{\partial^2}{\partial x^2} + \frac{n}{x^2}\frac{\partial}{\partial x} = \widehat{T}_1 + \widehat{T}_2$. Now,
\be \begin{split}
&\int (\widehat{T}_1\psi)^*\psi dx = \int \left(\frac{1}{x}\frac{\partial^2\psi}{\partial x^2}\right)^*\psi dx = \int\left(\frac{\partial^2\psi}{\partial x^2}\right)^*\frac{\psi}{x} dx = \left(\frac{\partial\psi}{\partial x}\right)^*{\frac{\psi}{x}}\Big{|}_b - \int \left(\frac{\partial\psi}{\partial x}\right)^*\left(\frac{1}{x}\frac{\partial\psi}{\partial x} - \frac{\psi}{x^2}\right)dx\\
& = -\psi^* \Big(\frac{1}{x}\frac{\partial\psi}{\partial x} - \frac{\psi}{x^2}\Big)\Big{|}_b + \int \psi^* \Big(\frac{1}{x}\frac{\partial^2\psi}{\partial x^2} - \frac{2}{x^2}\frac{\partial\psi}{\partial x} + \frac{2\psi}{x^3}\Big)dx = \int \psi^* \left(\frac{1}{x}\frac{\partial^2\psi}{\partial x^2} - \frac{2}{x^2}\frac{\partial\psi}{\partial x} + \frac{2\psi}{x^3}\right)dx
\end{split}\ee
Also
\be \begin{split}&\int (\widehat{T}_2\psi)^*\psi dx = \int \left(\frac{n}{x^2}\frac{\partial\psi}{\partial x}\right)^*\psi dx = n\int\left(\frac{\partial\psi}{\partial x}\right)^*\frac{\psi}{x^2} dx\\
& = n\psi^*{\frac{\psi}{x^2}}\Big{|}_b - n\int \psi^*\left(\frac{1}{x^2}\frac{\partial\psi}{\partial x} - \frac{2\psi}{x^3}\right)dx = -n\int \psi^* \left(\frac{1}{x^2}\frac{\partial\psi}{\partial x} - \frac{2\psi}{x^3}\right)dx
\end{split}\ee
In (B1) and (B2) we have dropped first terms appearing under integration by parts due to fall-off condition. Therefore,
\be \begin{split}
&\int (\widehat{T}\psi)^*\psi dx = \int [(\widehat{T}_1 + \widehat{T}_2)\psi]^*\psi dx = \int \psi^* \left(\frac{1}{x}\frac{\partial^2\psi}{\partial x^2} - \frac{(2+n)}{x^2}\frac{\partial\psi}{\partial x} + \frac{2(1+n)}{x^3}\psi\right)dx
\end{split}\ee
Now, for $n=-1$, $\widehat{T} = \frac{1}{x}\frac{\partial^2}{\partial x^2} - \frac{1}{x^2}\frac{\partial}{\partial x}$. Therefore
\be \begin{split}
\int (\widehat{T}\psi)^*\psi dx &= \int \psi^* \left(\frac{1}{x}\frac{\partial^2\psi}{\partial x^2} - \frac{1}{x^2}\frac{\partial\psi}{\partial x}\right)dx = \int \psi^*\widehat{T}\psi dx
\end{split}\ee
So, $\widehat{T}$ is hermitian for $n=-1$. Note that probability interpretation holds only for $n=-1$ also. The first term appearing in (75c) has been made hermitian by Weyl ordering. This may be proved as follows. Let
\be \widehat{T}_3 = i\hbar \left(2x {\partial \over\partial x} + 1\right)\ee
Therefore,
\be \int (\widehat{T}_3\psi)^*\psi dx = -i\hbar\int 2 x \left({\partial \psi\over\partial x}\right)^*\psi dx -i\hbar \int\psi^*\psi dx \ee
Now integrating first term by parts, we get
\be \int (\widehat{T}_3\psi)^*\psi dx =i\hbar\int\left(\psi^* x {\partial\psi\partial x} + \psi^*\psi\right)\ee
where, we have dropped the first term appearing under integration by parts due to fall-off condition. One can now clearly observe that
\be \int (\widehat{T}_3\psi)^*\psi dx = \int\psi^* \widehat{T}_3\psi dx\ee
and so the hermiticity of the second term has also been established. Therefore the effective Hamiltonian $\hat H_e$ is hermitian.\\

\end{document}